# A Novel Anticlustering Filtering Algorithm for the Prediction of Genes as a Drug Target

**Khalid Raza*, Akhilesh Mishra**

Department of Computer Science, Jamia Millia Islamia (Central University), New Delhi-110025, India
* kraza@jmi.ac.in

**Abstract**  The high-throughput data generated by microarray experiments provides complete set of genes being expressed in a given cell or in an organism under particular conditions. The analysis of these enormous data has opened a new dimension for the researchers. In this paper we describe a novel algorithm to microarray data analysis focusing on the identification of genes that are differentially expressed in particular internal or external conditions and which could be potential drug targets. The algorithm uses the time-series gene expression data as an input and recognizes genes which are expressed differentially. This algorithm implements standard statistics-based gene functional investigations, such as the log transformation, mean, log-sigmoid function, coefficient of variations, etc. It does not use clustering analysis. The proposed algorithm has been implemented in Perl. The time-series gene expression data on yeast *Saccharomyces cerevisiae* from the Stanford Microarray Database (SMD) consisting of 6154 genes have been taken for the validation of the algorithm. The developed method extracted 48 genes out of total 6154 genes. These genes are mostly responsible for the yeast's resistants at a high temperature.

## 1. Introduction

Microarray technology enables to measure the expression level of all or most of the genes in the genome simultaneously. The global scale gene expression profiling has revolutionized medical research allowing search for disease-related genes in a systematic and unbiased manner[1].

The identification of new genes from microarray data within a particular tissue type provide more reliable reference than are conventionally used techniques [2, 3-5]. Microarray data can be stratified on the basis of fold changes in expression [3], variance of expression [2, 5] or integrative correlations [4]. Candidate genes can be further selected from stratified data and frequently indicate expression stabilities [2-4]. However, the main requirement to the microarray data is an identification of new reference genes which demonstrate consistent stability across multiple tissue or cell types, and/or disease states [6].

As microarrays technologies have become more prevalent, the challenges associated with collection, management, and analysis of data from each experiment have essentially increased. Robust laboratory protocols improved understanding of the complex experimental design and decreased prices for some commercial platforms. These trends drive the field to more sophisticated experiments generating huge amounts of data [7]. With the help of these new technologies, we can find out answer of some challenging questions like (i) what are the functional roles of different genes and in what cellular processes do they participate?  (ii) how are genes regulated, how do genes and gene products interact, what are these interaction networks? (iii) how does gene expression level differ in various cell types and states, how is gene expression changed by various diseases and treatment?

In detail, the hybridized RNA is excited by a laser in microarray. The spot will be red, if the RNA from the sample population is in abundance. If the RNA from the control population is in abundance, it will be green. If sample and control bind equally, the spot will be appear yellow, while if neither binds, it will appear as black. Hence, from the fluorescence intensities and colors for each spot, the relative expression levels of the genes in the sample and control populations can be measured.

By measuring transcription levels of genes in an organism under different biological conditions, such as various developmental stages and diverse states, we can develop gene expression profiles that distinguish the dynamic functioning of each gene in the genome. The gene expression data are represented in the form of a table with rows indicating genes, columns representing samples (e.g. various developmental stages, tissues and diverse drug treatments). Each element of the matrix corresponds a number representing the expression level of the particular gene in the particular sample. We generally call such table as the gene expression matrix. For instance, if an over expression of certain genes is correlated with a certain disease, we can fix which conditions affect the expression of these genes and which other genes have similar expression profiles. Hence, we can investigate which compounds (potential drugs) lower the expression level of these genes [8].

DNA microarray array technology were used to investigate the functions of genes and also to the diagnosis of diseases [11]. With the help of microarray data analysis it was possible to discover drug targets [12]. In the recent times several methods have been developed for the analysis of gene expression data [13, 14]. Among these techniques, the clustering is most commonly used data analysis methods. Clustering generally groups the gene expression data with similar expression pattern, i.e. co-expressed genes. Due to various drawbacks of clustering techniques applied to gene expression analysis[15], we have proposed a novel algorithm which uses standard statistical functions, instead of clustering, for the analysis of differentially expressed genes. Our algorithm is also able to handle noises in the data, redundancy/replicate handling and testing the significance of data before analysing it. Finally, our algorithm extracts a list of genes which are differentially expressed in the dataset which can be used as the best drug target.

## 2. Proposed Algorithm

Our algorithm includes seven steps processing of data. At each step we eliminate some none useful gene or make the data more robust and systematic to help further processing of data. In this approach we have used very simple statistical technique to achieve the goal. These steps are:

**Step 1**. *Ratio and logarithmic conversion of microarray data*. When the raw fluorescence intensity (red) cy5 is plotted against cy3 (green) most of the data are clustered near the bottom left of the plot showing an asymmetric distribution of the raw data. This is thought to be result of imbalance of red and green intensities during plot sampling resulting in ineffective discrimination of differentially ex-pressed gene. For example, a gene that is up-regulated by a factor of 4 has an expression ratio of 4 (R/G = 4G/G = 4). However, for the case where gene is down-regulated by a factor of 4, the expression ratio becomes 0.25 (R/G = R/4R = ¼). Thus up-regulation is blown up and mapped between 1 and infinity, whereas down-regulation is compressed and mapped between 0 and 1.

$$upregulation \xrightarrow{mapped} [1, \infty]$$
$$downregulation \xrightarrow{mapped} [0, 1]$$

One way we use to improve data discrimination is to transform cy5 → cy3 value by taking the logarithm of base 2. The transformation produces more uniform distribution of data and has advantage to display up-regulated and down-regulated gene more symmetrically and more com-parable. To further normalize the data we put the data point horizontally by plotting the log ratio of cy5/cy3 against the average log intensities. In the representation the data are roughly symmetrically distributed around the horizontal axis. The differentially expressed gene then be more easily visualized. This form of representation is called '*intensity ratio plot*'. The linear regression is used in all these instances. A non-linear regression may produce a better fitting and help to eliminate the bias for data which not confirm to linear relationship owing to systematic sampling error. The most frequently used regression type is known as LOWESS (locally weighted scatter plot smoother) regression[9].

**Step 2.** *Elimination of gene that fail to provide data in majority of experiment*. In this step we remove that rows corresponding to gene that were not expressed or majority not expressed on any chip. In many of the cases due to some experimental problem some genes expression cannot be measured on the gene chip due to (i) wrong probing of gene on microarray chip, (ii) some specialized gene which are expressed in only a specific cell or specific condition are thus not expressed in that cell we are working on, (iii) scanner have some problem in that region to read the fluorescence value of gene, and (iv) due to defects in machine which make that microarray chip for Robotic probing. It is not hard and fast rule that if data is missing then we have to eliminate the row containing that gene. It can be a genuine problem that particular gene is actually not expressed in that particular condition.

In this algorithm we have considered that if missing values for a particular row are less than or equal to 40 % then missing values will be filled up by a zero value, indicating that genes are not expressed. If missing values in a row are more than 40 % then that particular row will be removed from the main dataset and will not be used further for analysis.

**Step 3.** *Analysis of significance of data.* In this step we check significance of data. The t-statistics is based on the assumption that the variability in these measurements follows a normal distribution, which means there is some pattern that is present in data which can be analyzed and may be interpreted as a result. Those data which are highly random and does not have any significance cannot be proceed for further analysis.

**Step 4.** *Replicate handling*. In replicate handling we remove those genes whose expression level are taken or noted more than one time in gene expression data. Thus, each gene should have only one entry. This will remove the redundancy in dataset. The multiple entry may produce due to presence of more than one position of single probe or different gene coding for same protein having different position or due to manual or machine error in detecting and noting expression level of gene. These redundancies will increase the volume of data as well as analysis time. This step is optional if we are sure that our data is quite mature and it does not have redundancy.

**Step 5.** *Elimination of gene having less than two-fold change in expression level*. We eliminated those genes that do not show considerable variation in expression level. In the dataset, positive value means up-regulation of expression in cy5 labeled gene and negative value means down-regulation of cy5 labeled gene. Those genes which neither show up-regulation nor down-regulation at least of half of its normal condition or which have less variation in expression level in control and diseased condition are not useful [10].

Thus, we have filtered the data and taken only those genes which show variation in expression level more than half of its expression level in control condition. For this, we have taken mean of each row and extracted only those rows or gene which have 1≤mean<−1, that actually represent change in expression level of at least half of its normal condition.

**Step 6.** *Conversion of datasets using logsigmoid function.* At this step we have used log-sigmoid function to transform values in the range [0, 1] to make data more convergent which help us for further analysis. This conversion function is also useful as it transform all the negative values in positive range which is very useful for statistical data analysis. The log-sigmoid transformation takes the input, which can have any value between [+∞ to −∞] and squashes the output into the range [0, 1]. The transfer function is given by,

$$\log sigmoid(x) = \frac{1}{1+e^{-x}}$$

**Step 7.** *Elimination of genes that have high variation across the collection of sample.* At this step we remove those genes which do not have consistent variation or variance in expression level in all different experimental condition in different time series. In this algorithm, we have eliminated those genes which have more than 36 % of variation because we are interested in those genes which show consistent differential expression level in disease case. Thus, we can use that gene or gene product for drug target to inhibit the symptom of that particular disease case.

Suppose we have n number of genes at m different time points, expression level of gene n will be $x_{n,1}, x_{n,2}, x_{n,3}, ...., x_{n,m}$. We have calculated the coefficient of variance (CV) for each row in the dataset. The CV is given by,

$$C.V. = \frac{SD}{\bar{x}} \times 100$$

where SD and $\bar{x}$ are standard deviation and mean respectively. By using coefficient of variance those genes which show less than 36 % variation are selected as they show consistency and other that show more than 36 % variation are deleted out because they are not representing as marker gene of that state. This cut off can be changed according to our need, bigger the cut off bigger will be the output. The flow chart of the proposed algorithm is presented in Figure 1.

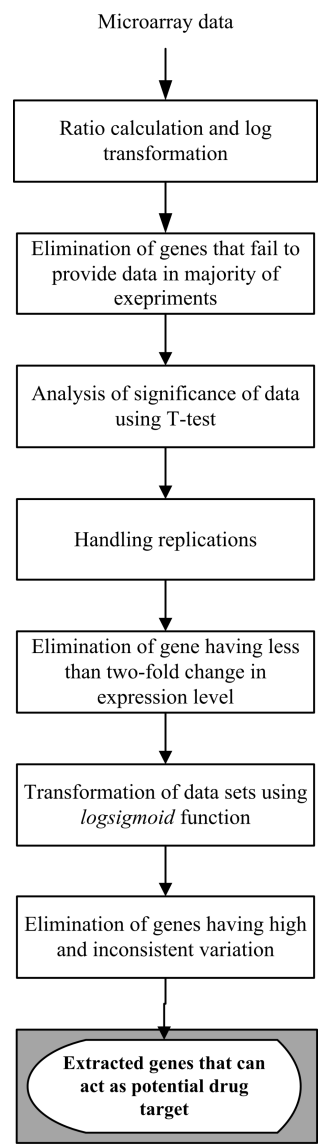

**Figure 1.** Flow chart of the proposed algorithm

## 3. Evaluation of the Algorithm

The proposed algorithm has been implemented in Perl programming language. We have tested the proposed algorithm on time-series gene expression data of Saccharomyces cerevisiae (yeast) consisting of 6,154 genes and found that our algorithm extracts only 48 genes, i.e. 0.78 % of total number of genes. Thus, the proposed algorithm is highly efficient as it extract only a small number of genes as the candidate genes from whole dataset. Its performance in respect of time and space complexity is moderately well. The asymptotic time complexity of the algorithm is the quadratic time **$O(n^2)$**. The main beauty of the proposed algorithm is that it is quite simple which can be easily implemented by anyone using almost any programming language. As we know there is no consistent data format in microarray technology, so we can easily adjust the algorithm parameters to make it work for almost any file format. The other main feature of our algorithm is no use of clustering approach. Clustering approach suffers from (i) the problem of subjective nature of deciding which external criteria to use in the absence of formal statistical tests, to choose the number of clusters, (ii) it is also guaranteed to produce clusters from any data and there is currently no generally accepted way to test a null hypothesis of no clusters (e.g., data are distributed uniformly). For these reason, caution is required in interpreting the results of a cluster analysis method. The results always need to be examined to see if it is plausible that they are indeed natural clusters and not just artifacts of the algorithm[15].

## 4. Results and Discussions

We have tested the proposed algorithm on the time-series gene expression data of Saccharomyces cerevisiae from Stanford Microarray Database (SMD) consisting of 6,154 genes[16]. The program filtered 48 genes out of 6154. These are those genes which are majorly responsible for making yeast cell more resistant and tolerable at high temperature. Filtered genes with their coefficient of variance (CV) are given in Table 1. After analyzing the result it is found that out of 48 extracted genes, 32 genes are directly responsible for providing resistance tolerance for high temperature. In other view, if we suppress some selected gene from them and make the cell more susceptible for heat, then a small raise in temperature can kill the cell. Thus, these genes can act as best candidate for drug target in Saccharomyces species.

**Table 1.** A list of filtered genes with their *CV* values

| Extracted Genes | C.V. | Extracted Genes | C.V. |
|---|---|---|---|
| YAL060W | 35.73 | **YIL136W** | 35.62 |
| YBR001C | 35.84 | YKL103C | 35.89 |
| YBR072W | 35.80 | YLL026W | 35.40 |
| YBR169C | 35.58 | YLL039C | 35.89 |
| **YCL042W** | 35.87 | YLR178C | 35.51 |
| YDL021W | 35.89 | YLR216C | 35.58 |
| YDL022W | 35.68 | YLR251W | 35.98 |
| **YDL023C** | 35.80 | YLR258W | 35.96 |
| YDL124W | 35.93 | YLR259C | 35.81 |
| YDR001C | 35.95 | **YLR327C** | 35.45 |
| YDR074W | 35.87 | YML100W | 35.37 |
| YDR171W | 35.39 | **YML128C** | 35.59 |
| YDR214W | 35.39 | YMR105C | 35.54 |
| YDR258C | 35.37 | YMR186W | 35.74 |
| YER020W | 35.90 | **YMR251W-A** | 35.88 |
| YER103W | 35.88 | YNL007C | 35.97 |
| YFL014W | 35.36 | YNL015W | 35.92 |
| YGL037C | 35.64 | YNL274C | 35.92 |
| YGR142W | 35.50 | YOL117W | 35.74 |
| YGR146C | 35.56 | YOR020C | 35.57 |
| YGR248W | 35.44 | YOR027W | 35.67 |
| **YHL021C** | 35.83 | **YOR052C** | 35.77 |
| YHR104W | 35.50 | YOR230W | 35.91 |
| YIL111W | 35.62 | YPL240C | 35.75 |

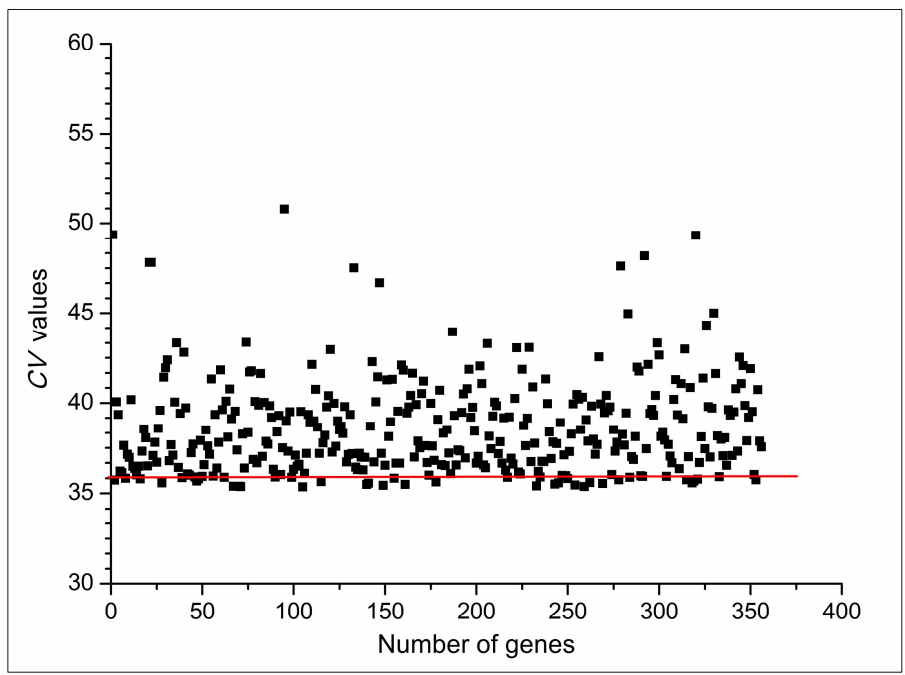

**Figure 2.** Dot plot of *C.V.* values

In addition, the result also provide that 8 genes (namely, YIL136W, YLR327C, YML128C, YMR251W-A, YOR052C, YCL042W, YDL023C and YHL021C) with unknown function that show more than two-fold change in expression level at high temperature. Thus, these can be better target or can be a novel field of research to find the function and participation of these genes in heat tolerance. Further, these 8 genes which are indirectly involved in heat tolerance like by harvesting energy, involving in cell signalling etc, whose role is not yet correlated with heat tolerance but it can be find out and this can also be a better drug target. A distribution of genes and their *CV* values (in %) are represented in the form of dot-plot in Figure 2. This figure has been plotted on the basis of intermediate results which have been obtained after the step 6 of the algorithm. At the step 6, we have got around 356 genes whose distributions are shown in the dot-plot. The thick-line in the Figure 2 indicates our threshold value (CV <= 36 %).

## 5. Conclusions

Microarray is a high throughput technique to analyze the gene expression of particular cell at the specific conditions. This high throughput data can further be utilized to identify the gene as a drug target. We proposed an anticlustering algorithm by using some standard statistical methods which is able to identify and filter genes as a drug target. Our algorithm allows to find a wide range of genes that 67 % of genes are directly involved in particular events. In addition, 17 % genes having unknown function and 17 % genes are indirectly related with diseases are also identified. So this algorithm can be used to investigate genes for particular disease and it also opens a new dimension of research to find the function of genes whose functions are still unknown and to find the role of genes which are indirectly involved in disease state. The algorithm presented here is a fast and scalable for the identification and filtration of genes as a drug target.